# Improving electron tomography of mesoporous silica by Ga intrusion

Alexander Kichigin[a], Johannes Böhmer[a], Moritz Buwen[a], Benjamin Apeleo Zubiri[a], Mingjian Wu[a], Johannes Will[a], Dominik Drobek[a], Alexander Götz[a], Nora Vorlaufer[c], Jakob Söllner[b], Matthias Thommes[b], Peter Felfer[c], Thomas Przybilla[a]*, Erdmann Spiecker[a]*

[a] Institute of Micro- and Nanostructure Research (IMN) and Center for Nanoanalysis and Electron Microscopy (CENEM), Interdisciplinary Center for Nanostructured Films (IZNF), Department of Materials Science, FAU Erlangen-Nuremberg, Germany

[b] Institute of Separation Science and Technology, Department of Chemical and Biological Engineering, FAU Erlangen-Nuremberg, Germany

[c] Institute for General Materials Properties, Department of Materials Science, FAU Erlangen-Nuremberg, Germany

*Corresponding authors: thomas.przybilla@fau.de, erdmann.spiecker@fau.de

**Abstract:**

Electron tomography (ET) offers nanoscale 3D characterization of mesoporous materials but is often limited by their low scattering contrast. Here, we introduce a gallium (Ga) intrusion strategy for mesoporous silica that dramatically improves imaging contrast – a key benefit that enables more accurate 3D reconstructions. By infiltrating Ga through a modified mercury intrusion porosimetry process, the high-angle annular dark-field (HAADF) STEM signal is enhanced by 5 times, resulting in a 34% improvement in reconstruction resolution and a 49% enhancement in interface sharpness. In addition, the increased sample conductivity facilitates focused ion beam (FIB) milling by minimizing charging effects and reducing drift. This approach enables precise segmentation and quantitative analysis of pore connectivity and size distribution, thereby extending the applicability of ET to light-element non-conductive materials and advancing structure-property characterization of complex porous systems.

**Graphical abstract:**

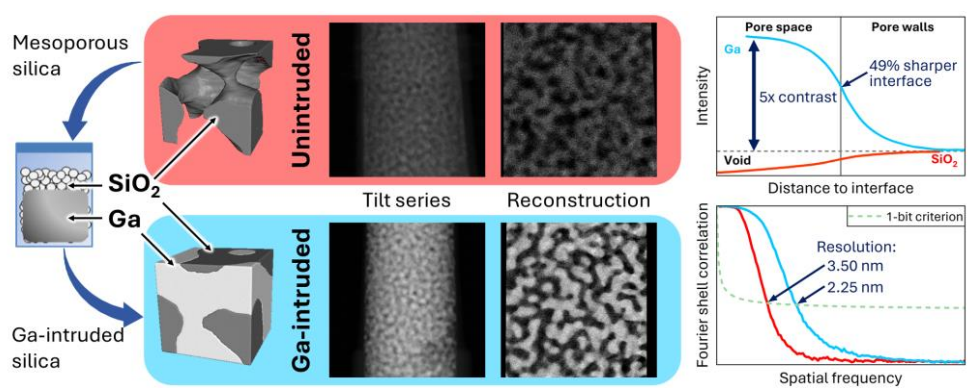

## 1. Introduction

The growing demand for porous materials across diverse technical applications has spurred the development of increasingly sophisticated characterization techniques. Among these materials, metal oxides (e.g., titanium oxide, hematite, zinc oxide), zeolites, and porous silica are particularly valued for their chemical stability, large surface area, uniform pore structure, and precisely tuneable properties [1,2]. These attributes make them well suited for use in drug delivery [3], catalysis [4], adsorption [5], as well as energy storage and conversion [6]. Achieving optimal performance in each of these domains requires accurate pore network characterization, as it is crucial for fine-tuning of structural features to enhance efficiency, durability, and overall functionality.

A variety of methods exists to investigate sub-micrometer pore networks, each offering unique advantages and constraints. Physisorption (e.g. Nitrogen at 77 K and Argon at 87 K) and mercury porosimetry are widely employed techniques for obtaining key textural properties, yet they do not offer direct structural images. Advanced physisorption methodologies, particularly when combined with statistical mechanics-based approaches such as non-local density functional theory (NLDFT) and grand canonical Monte Carlo (GCMC) simulations, are well-suited for providing reliable pore size and volume information, surface area measurements, and insights into key pore network characteristics within mesoporous structures [7–9]. Small-angle X-ray scattering (SAXS) [10] is a non-destructive technique for bulk sample analysis, yet it does not provide spatially resolved information on the pore network. Electron tomography (ET) [11] is a non-destructive technique that enables three-dimensional (3D) reconstructions of porous architectures at nanoscale resolution. Furthermore, focused ion beam (FIB) scanning electron microscopy (SEM) tomography [12] can achieve imaging and depth resolutions down to a few nanometers but is inherently destructive, as it involves sequential milling of layers to expose internal structures.

Pores are typically classified as micropores (< 2 nm), mesopores (2–50 nm), and macropores (> 50 nm) [13], and many materials utilize hierarchical pore networks that integrate these pore sizes to enhance functionalities such as mass transport, catalytic activity, or adsorption capacity. In this study, we chose mesoporous silica due to its well-defined mesopore structure, which balances large surface area with efficient diffusion pathways – properties that are particularly valuable in applications like catalysis and chromatography [4]. To examine these mesoporous materials at the nanometer scale in three dimensions, 360° HAADF-STEM electron tomography (360°-ET) has emerged as a powerful approach. By capturing tilt angles over a full 360-degree range, it enables more complete reconstructions by mitigating the missing-wedge problem associated with conventional ET [14]. The missing-wedge problem refers to the incomplete angular sampling in conventional ET, which can lead to anisotropic resolution and reconstruction artifacts. 360°-ET allows achieving more accurate and isotropic 3D reconstructions, enhancing our ability to comprehensively analyse pore structures.

Despite its potential for high-resolution 3D imaging, 360°-ET of mesoporous silica still faces several challenges. The HAADF signal intensity depends on the atomic number (Z) approximately as $Z^n$ where $n$ lies in the range between 1.7 and 2.0 depending on the detector [15], therefore light-element materials like silica ($Z_{SiO_2} \approx 10$) exhibit poor contrast. This issue is made worse due to carbon contamination ($Z_C = 6$) [16], which can obscure pore structures when it accumulates on the sample. Additionally, beam-induced damage remains a persistent challenge for ET of materials with low mechanical stability [17]. Beam damage refers to deformations in the porous network caused by radiolytic and knock-on damage from the high-energy electron beam used during ET. Due to the delicate nature of mesoporous silica, intense electron irradiation can weaken or even break the pore walls, leading to structural distortions such as thinning or collapse [18,19]. These modifications compromise the quality of the acquired tilt series data, ultimately affecting the reliability of 3D reconstructions.

Sample preparation also presents challenges, especially with non-conductive materials such as silica. FIB milling is typically used to create thin pillars for 360°-ET [20], but the ion beam induces charging on the silica surface. This charge buildup distorts the ion beam's trajectory, resulting in uncontrolled material removal and more severe curtaining – a common artifact characterized by vertical lines or grooves on the milled surface [21]. These distortions arise due to variations in sputtering rates caused by localized charging and can lead to imprecise milling, defects in sample geometry, and a roughened surface, ultimately compromising the quality of the prepared sample. Simultaneously, charging during SEM imaging degrades both imaging stability and resolution. Together with the charging-induced artifacts during FIB milling these issues lead to milled samples with structural imperfections. As a result, the fidelity of the 3D reconstructions is significantly reduced, undermining the accuracy of the subsequent structural analyses.

In this study, we aim to achieve precise, quantitative 3D analysis of a light-element porous material by enhancing the contrast in 360°-ET through gallium intrusion (Ga, $Z_{Ga} = 31$) into the mesoporous silica pore structure (see Figure 1 for a schematic overview). This approach is analogous to the "staining" technique used in medical imaging, where contrast is enhanced by introducing a contrasting medium [22]. The Ga intrusion is performed using a modified mercury intrusion porosimetry (MIP) process [23], which has been adapted to facilitate the infiltration of Ga into the pore network under controlled pressure conditions. By adjusting the applied pressure, we can tune the fraction of pores that become infiltrated with Ga. A key advantage of Ga lies in its ability to wet the pore walls and remain within the pore network after the pressure is released, in contrast to mercury, which flows out upon depressurization. Furthermore, Ga is less toxic than mercury, which eases sample handling and contributes to laboratory safety. Additionally, Ga intrusion mitigates charging effects during FIB milling, thereby simplifying sample preparation. Ultimately, this method allows us to overcome the low contrast of silica and achieve more reliable 3D reconstructions, thereby facilitating a more thorough structure-property analysis of mesoporous materials. This advancement not only enhances our understanding of mesoporous silica but also extends the applicability of ET-based characterization to a broader range of light-element non-conductive porous materials.

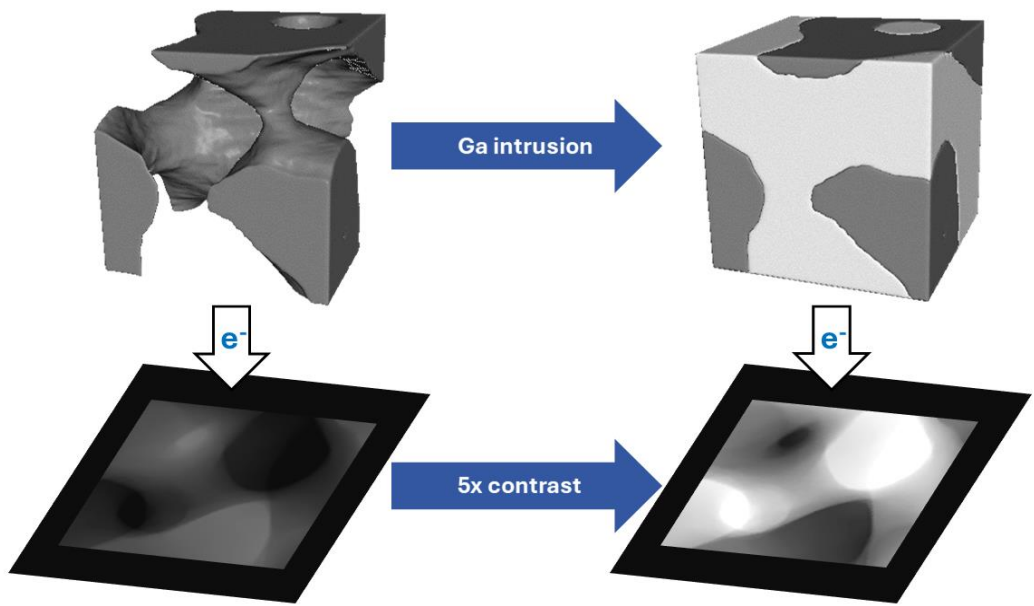

*Figure 1. Schematic representation illustrating the enhancement of 3D reconstruction contrast in mesoporous silica achieved through Ga intrusion. The left panel shows the low-contrast porous structure before Ga intrusion, while the right panel demonstrates the improved contrast after the infiltration of Ga into the pore network.*

## 2. Materials and methods

### 2.1. Experimental workflow

The experimental workflow consists of three main stages: Ga intrusion of the mesoporous silica network (Figure 2a), FIB pillar preparation (Figure 2b), and the acquisition of a 360°-ET tilt series (Figure 2c). This sequence provides the necessary data for subsequent quantitative 3D reconstructions and analyses.

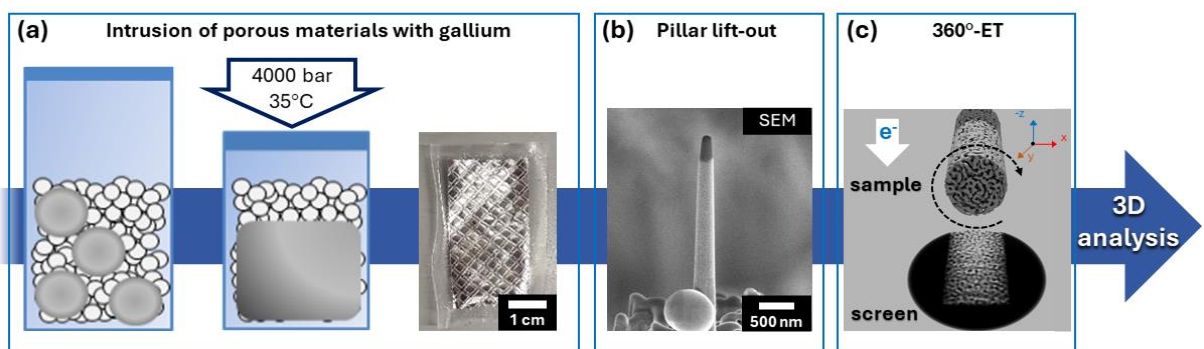

*Figure 2: Overview of the experimental workflow for investigating mesoporous silica using Ga intrusion and 360° electron tomography. (a) Mesoporous silica is intruded with Ga under controlled pressure (4000 bar) and temperature (35°C) conditions. (b) A pillar-shaped sample is prepared via focused ion beam milling for electron tomography. (c) A 360° electron tomography experiment is performed. Then the data undergoes alignment, reconstruction, and segmentation for detailed structural analysis.*

### 2.2. Intrusion of the CPG with Ga

Intrusion of the controlled pore glass (CPG) with Ga was performed using a modified setup based on a commercial mercury porosimeter which was placed in a thermostated housing. The pure material consisted of mesoporous silica particles with a width of approximately 150 μm. The corresponding procedure has been performed in a way similar to a previous publication [24] with the difference that Ga instead of water was used in this work. The investigated controlled pore glass is a (discontinued) reference material available from the "Bundesanstalt für Materialforschung und -prüfung" (BAM, Berlin, Germany) which was distributed as FD121.

In this work, FD121 samples were subjected to three different levels of Ga intrusion for comparative analysis: 0% (unintruded), 100%-intruded, and 50%-intruded. The sample labels, summarized in Table 1, consist of a letter indicating the intrusion level (e.g., U for unintruded, I for 100%-intruded, and H for half intruded) followed by a number representing the sequential order in which the samples were prepared. This numbering reflects the iterative adjustments made to the experimental parameters to achieve successful 360°-ET measurements.

*Table 1. Summary of sample names and Ga intrusion levels*

| Sample name | Ga intrusion level | Description |
|---|---|---|
| **U11** | 0% | Unintruded |
| **I07** | 100% | Fully intruded |
| **H02** | 50% | Half intruded |

*2.3. FIB pillar preparation*

Pillar-shaped specimens were prepared for 360°-ET using a FIB-SEM Helios NanoLab 660 DualBeam system. Before FIB milling, the unintruded sample U11 needed more elaborate pre-preparation steps than the intruded samples. One particle was carefully glued onto the top of an electron tomography needle made of a copper-beryllium alloy using silver glue (Acheson Silver DAG 1415). Before the glue dried, the particle's position was precisely adjusted to ensure it was centred at the needle tip, and the alignment was verified under a stereo light microscope.

To reduce milling time in the FIB, the unintruded silica particle was pre-cut into a pillar shape with a 50 μm diameter (see Figure S5) via laser ablation using a 3D-Micromac microPREP system. The sample was then coated with a thin carbon layer (25 nm) using a Leica EM ACE200 sputter coater to minimize sample drift and charging during subsequent FIB cutting. Additionally, a charge neutralizer was activated inside the FIB-SEM device to further reduce charging effects during the milling procedure with Ga ions. Once the pillar reached the diameter of 5 μm, it was coated again with a 25 nm thick carbon layer in the carbon coater before being thinned to a final diameter of approximately 250 nm and a length of 1-2 μm via FIB milling. The final diameter of 250 nm was chosen to make the sample thin enough to minimize the beam broadening in STEM [25], while maintaining a sufficient number of pores within the volume for data analysis.

For the 100%-intruded (I07) and 50%-intruded (H02) samples, pillars were prepared using the standard lift-out procedure [26]. Each sample was glued onto a carbon stub using the same silver glue. The region of interest was coated with carbon using the gas injection system (GIS) inside the FIB. A square block of approximately 2 μm in side-length and 5 μm in depth was cut from the material with the FIB and attached to the manipulator (Thermo Scientific EasyLift Nanomanipulator). The block was then transferred to the electron tomography needle, secured with carbon using the gas injection system, and further thinned down to a final diameter of around 250 nm. The final cutting step for all samples was completed in a single pass using an ion beam current of 7.7 pA at 30 kV.

*2.4. Electron tomography*

To characterize the internal structure of the pillars, 360°-ET was performed in a double-Cs-corrected FEI Titan³ Themis TEM operating at 300 kV and equipped with a Fischione Model 2050 On-Axis Rotation Tomography holder. To minimize beam-induced deformation of the unintruded silica (Figure S6) and mitigate Ga depletion from the intruded silica (Figure S7), tilt series images were acquired under the same low-dose conditions: beam current of 9 pA, dwell time of 5 µs, and pixel size of 0.27 nm, resulting in an electron dose of 39 e⁻/Å² per tilt. Auto-focusing was performed on a separate area adjacent to the imaging region. The convergence semi-angle was set to 3 mrad for all pillars, by using a custom 10 µm probe-forming condenser aperture in aberration-corrected nanoprobe mode. Due to the sensitivity of the samples, conventional contamination-mitigating techniques (baking, plasma cleaning, beam showering) could not be employed. Instead, each sample remained in the microscope column overnight prior to data acquisition to stabilize and reduce contamination probability.

Table 2 summarizes the 360°-ET imaging parameters used for each pillar. For the unintruded (U11) and fully intruded (I07) samples, we employed a default "balanced" HAADF-STEM camera length (CL) of 91 mm (collection angle: 61–200 mrad). This CL setting is sufficient for most materials, particularly when they are amorphous or only weakly crystalline, as in the case of this work. Using the same parameters for U11 and I07 ensured a direct, quantitative comparison of their reconstruction resolutions.

In contrast, for the 50%-intruded (H02) sample, we employed a higher CL of 185 mm (collection angle: 30–184 mrad) to explore whether increased signal could be achieved with the same electron dose – one other key advantage for imaging amorphous light-element beam-sensitive materials. On the other hand, in the case of strongly crystalline materials, the CL should be reduced to minimize Bragg diffraction artifacts, but since our samples are amorphous, Bragg effects are not a concern.

*Table 2. Image acquisition parameters for 360°-ET tilt series of the pillars.*

|  | **U11** | **I07** | **H02** |
| --- | --- | --- | --- |
| **Camera length, mm** | 91 | 91 | 185 |
| **Collection angle (HAADF), mrad** | 61-200 | 61-200 | 30-184 |
| **Tilt step, °** | 0.5 | 0.5 | 1 |
| **Number of projections** | 360 | 360 | 180 |

*2.5. Data analysis*

The tilt series were reconstructed using a quantitative, standardized approach to ensure comparability of results across all samples. Before each tomography experiment, an image of the HAADF-STEM detector (Figure S8) was recorded under the gain settings chosen for the experiment, providing the maximum electron count signal. After the acquisition, the tilt series signal $I_{tilt}$ was normalized by dividing it by the average detector area signal $I_{det}$, subtracting the noise level $I_{min}$ first. This process yielded the quantified signal $I_q$ [27], calculated as

$$I_q = \frac{I_{tilt} - I_{min}}{I_{det} - I_{min}}. \tag{1}$$

Prior to reconstruction, the tilt series were pre-processed using a robust stack alignment routine based on the center-of-mass projections algorithm [28] implemented in the "ToReAl" Matlab script. This step efficiently corrected for sample shifts, ensuring that all projections were accurately registered and thus minimizing artifacts in the final reconstructions. Next, a three-dimensional reconstruction was performed using the Simultaneous Iterative Reconstruction Technique (SIRT) [29–31] with 100 iterations.

To evaluate the resolution of the reconstructed volumes, the Fourier shell correlation (FSC) [32,33] was calculated for a selected cubic region within each reconstructed 3D dataset. In this process, following the procedure described in [33], the tilt series were divided into even and odd frames, and each half was reconstructed using the same parameters as the full dataset. The resulting sub-reconstructions were Fourier transformed, and the FSC correlation function was plotted as a function of spatial frequency.

The 1-bit and ½-bit threshold curves were then superimposed on the same graph. These thresholds serve as standard benchmarks: the 1-bit criterion indicates the spatial frequency at which the data carries about one bit of information per voxel (roughly when the signal-to-noise ratio is unity), while the ½-bit criterion is less conservative and typically gives a better resolution estimate. The intersection points of the FSC curve with these threshold curves indicate the corresponding resolution limits.

For segmentation and 3D visualization, Arivis Vision4D (Version 4.1.0, Carl Zeiss Microscopy Deutschland GmbH) was used. The software's machine-learning segmentation (based on random forest algorithm [34]) considered only local voxel neighbourhoods. To further analyse pore architecture, the segmented data were imported into Avizo (Version 2022.2, Thermo Fisher Scientific Inc.) and cropped to $600 \times 600 \times 1000$ pixels. A watershed algorithm was applied, and pores were separated from their neighbours (Figure S8c). The pore network was then skeletonized for connectivity and tortuosity analysis (Figure S8d,e). Tortuosity was assessed by simulating fluid intrusion in the z-direction (longest axis), ignoring throats with an equivalent radius below four voxels (1.08 nm). Connectivity was evaluated by counting attached throats or neighbouring pores. In simple terms, tortuosity measures how winding or indirect the paths through the pores are compared to a straight line, while connectivity refers to how many neighbouring pores each pore is linked to.

A second quantitative method based on proximity histogram (proxigram) analysis was employed to evaluate interface sharpness. Traditionally applied to atom probe tomography (APT) data to show atomic concentration gradients near interfaces, proxigrams can also display voxel-intensity variations. In this study we modified existing APT Matlab code [35] to analyse an ET dataset by examining how voxel intensities vary with distance from a defined interface. As illustrated in Figure 3, a synthetic grayscale dataset (Figure 3a) was first segmented into two phases (Figure 3b). Using the voxel coordinates and intensity values from the reconstructed volume, along with the isosurface (defining the interface) from the segmented data, we computed the shortest distance from each voxel to the interface (Figure 3c). Voxel intensities were then plotted against these distances (with negative and positive signs assigned to the two phases) and averaged for voxels at the same distance to produce a normalized intensity profile (Figure 3d). Finally, the spatial distance between the 25% and 75% intensity thresholds (the $r_{25-75}$ factor [25]) was used to quantify interface sharpness, where smaller values indicate a sharper transition, and larger values indicate a more diffuse interface.

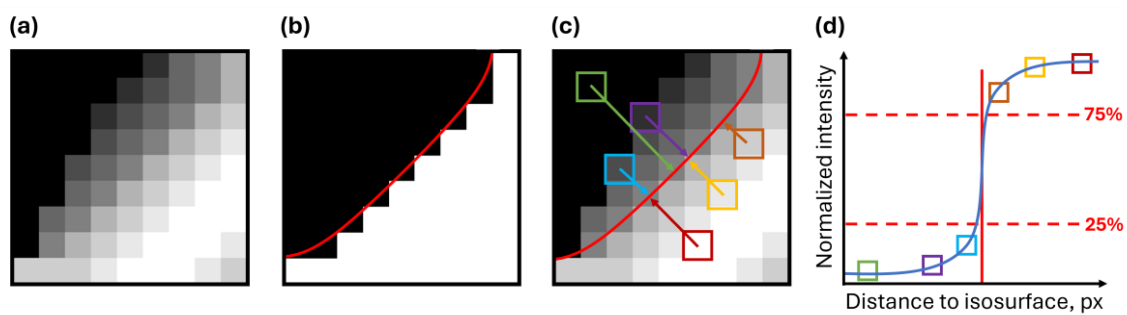

*Figure 3. Scheme of the proxigram evaluation workflow. (a) An arbitrary grayscale image (2D slice) from the 3D reconstruction. (b) The corresponding user-defined binary segmentation*

*including the determined isosurface indicated with a red line. (c) The superposition of the grayscale and binary images, where coloured squares illustrate specific pixels from (a) and their distances to the isosurface from (b). (d) A conceptual depiction of the resulting proxigram, showing intensity changes as a function of distance from the interface.*

Finally, the pore size distribution (PSD) was computed from the segmented 3D reconstructions. In this context, local thickness (representing the pore size at any given point) is defined as the diameter of the largest sphere that can be inscribed at that location while remaining entirely within the structure, a method known as Maximum Sphere Inscription (MSI) [36]. To speed up the calculation, a Euclidean distance map was generated. This map calculates the shortest distance from each voxel to the nearest surface, and this distance corresponds to the radius of the largest sphere that can fit at that voxel's location. These distance values are then used to virtually fill the entire 3D volume with spheres of various diameters. By counting the number of voxels associated with each sphere size, the PSD is obtained. The fast local thickness algorithm [37] was used in this work because of its lower computational complexity: $\mathcal{O}(x^4)$ compared to $\mathcal{O}(x^7)$ in the case of conventional local thickness algorithm. This efficiency is particularly beneficial for unintruded silica, where the background has the same reconstructed intensity as the pores, greatly increasing the data volume for the PSD evaluation.

## 3. Results

### 3.1. Ga intrusion mitigates charging and improves sample preparation

Backscattered electron (BSE) images of the bulk mesoporous silica (Figure 4a–c) clearly illustrate the benefits of Ga intrusion. The unintruded sample exhibits severe charging and significant image shifts during SEM imaging. In contrast, the fully intruded sample exhibits no appreciable drift, and even the 50%-intruded sample was stable during imaging. These observations confirm that Ga intrusion introduces sufficient conductivity to the originally non-conductive silica, thereby facilitating FIB milling. Notably, U11 required additional laser pre-cutting and extra charge-mitigation steps during pillar preparation, while I07 and H02 could be processed using standard lift-out protocols.

BSE images also reveal notable variations in Ga infiltration between samples. The 100%-intruded sample exhibits predominantly homogeneous intrusion, although isolated unintruded pores are still evident. These voids are likely associated with minor localized variations in pore size or the presence of trapped gas pockets within the structure. In contrast, in the 50%-intruded sample Ga is not uniformly distributed; instead, it condenses within certain pores while leaving others empty. This heterogeneous intrusion likely results from insufficient wetting of the pore walls, which may be influenced by specific surface chemical interactions.

### 3.2. Enhanced contrast in tomography tilt series

Representative HAADF-STEM images from the 360° tilt series (Figure 4d–f and Supporting Video S1) further underscore the advantage of Ga intrusion. Although all specimens suffer from some degree of carbon contamination, the unintruded U11 shows low contrast between silica and contamination due to their similar low atomic numbers. In contrast, both I07 and H02 display a much brighter signal in the Ga-intruded pore regions. Notably, the half-intruded H02 sample, for the interest of future research, was acquired at a longer camera length (185 mm instead of the default 91 mm), which for the Ga phase yielded ~56% higher HAADF signal under the same electron dose (Figure S8).

## 3.3. Reconstruction and segmentation quality

Slices from the reconstructed tomograms (Figure 4g–i and Supporting Video S2) reveal that U11 suffers from low grey-level differentiation between silica and voids, which complicates further segmentation. By contrast, the Ga-intruded samples (I07 and H02) show clearly isolated grey values between the Ga-intruded pores and the silica matrix, allowing more straightforward segmentation (Figure 4j–l and Supporting Video S3). In U11, a simple two-phase segmentation (silica vs. pores/background) was sufficient. For the fully intruded I07 sample, segmentation isolated the Ga phase with silica and background appearing similarly dark, while in H02 three phases (silica, Ga, and unintruded pores) could be reliably distinguished.

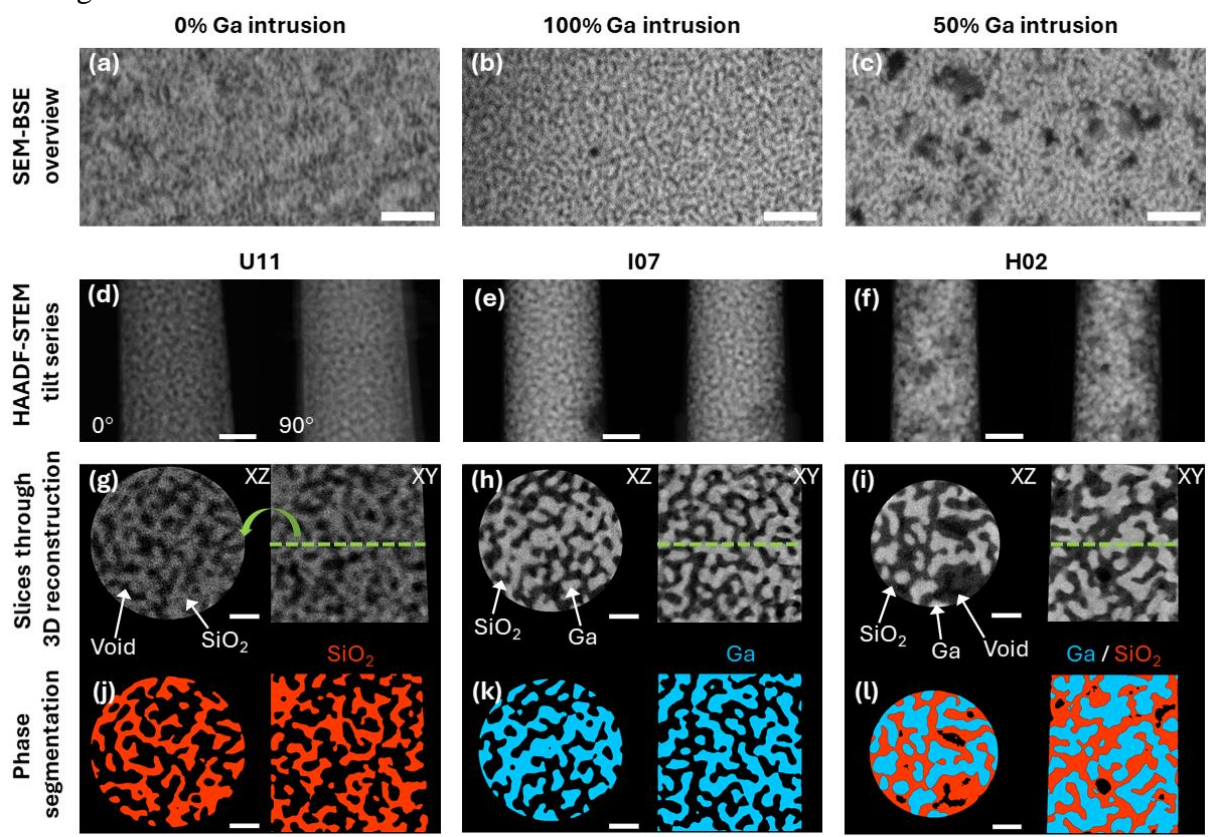

*Figure 4. Overview of the imaging and analysis data for the unintruded (U11), 100%-intruded (I07), and 50%-intruded (H02) mesoporous silica samples. (a–c) SEM-BSE images highlighting differences in behaviour at various intrusion states: the unintruded silica exhibits significant charging and image shifts, while the 100%- and 50%-intruded demonstrate stable imaging due to Ga enhanced conductivity. Scale bar corresponds to 200 nm. (d–f) Representative HAADF-STEM tilt series images acquired at 0° and 90°, showing variations in contrast and imaging quality. Scale bar corresponds to 100 nm. (g–i) 2D slices from the reconstructed ET volumes (XZ and XY slices), demonstrating improved contrast in the intruded samples, which facilitates segmentation. Scale bar corresponds to 50 nm. (j–l) Segmentation of the same reconstructed slices, showcasing enhanced delineation of pore structures in the intruded samples. In the H02 sample, all three phases (silica, Ga, and unintruded pores) are distinguished. Scale bar corresponds to 50 nm.*

## 3.4. Quantitative analysis of interface sharpness and resolution

Quantitative analysis using proxigram evaluation (Figure 5a) reveals that the HAADF intensity from Ga in I07 is 5.5 times higher than that from silica, which leads to 5 times increase in contrast compared to U11. Moreover, the interface sharpness (as determined by the $r_{25-75}$ parameter) improves from 4.7 nm in U11 to 2.4 nm in I07. FSC analysis (Figure 5b) confirms that the overall resolution is enhanced from 3.5 nm (U11) to 2.3 nm (I07) at the 1-bit criterion, corresponding to a 34% improvement (see Table 3).

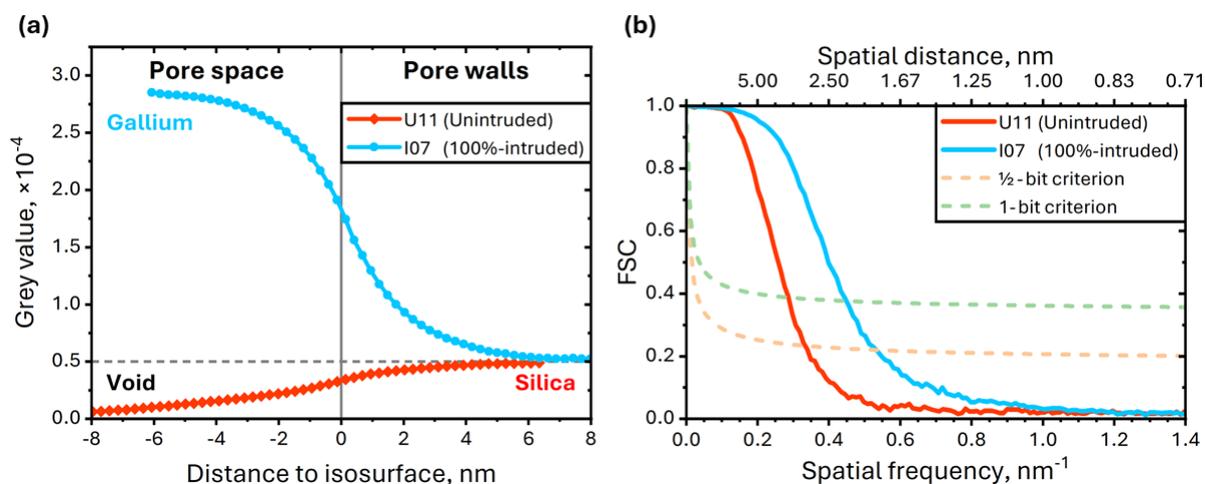

*Figure 5. Quantitative analysis of the resolution improvement. (a) Results of proxigram evaluation. The grey value corresponds to quantitative mass-thickness intensity in the reconstructed volume. (b) Fourier shell correlation analysis.*

The enhancement in HAADF-STEM contrast is further validated by theoretical predictions (see S2 in the Supporting Information for more details) based on effective atomic number scaling ($I_{HAADF} = Z_{eff}^n$) and molar mass relation. With effective atomic numbers of approximately 10 for $SiO_2$ and 31 for Ga, and fitting an exponent n ≈ 1.8, the model predicts a contrast enhancement of about 5.1 times. This is in excellent agreement with the experimental observation of a 5 times improvement, confirming the effectiveness of Ga intrusion for boosting the signal in light-element beam-sensitive materials.

*Table 3. Summary of resolution evaluation with FSC and proxigram analysis for unintruded and 100%-intruded samples*

| Resolution | U11, nm | I07, nm | Improvement |
|---|---|---|---|
| **FSC 1-bit** | 3.5 | 2.3 | 34% |
| **FSC ½-bit** | 3.0 | 1.9 | 37% |
| **Interface sharpness** | 4.7 | 2.4 | 49% |

*3.5. Pore network and size distribution analysis*

Segmentation of the reconstructed volumes allowed for a detailed analysis of the pore network. Both U11 and I07 exhibit similar intrinsic network properties: tortuosity is 1.57 in both cases, and the pore connectivity averages to a value of 6.46 and 6.38 (Figure S11), respectively.

Importantly, the pore size distribution (PSD) determined using the Fast local thickness algorithm reveals a mean pore size of 17.8 ± 3.0 nm in U11 vs 15.9 ± 2.8 nm in I07 (Figure 6 and Supporting Video S4), these values align closely with the 18.8 nm mean obtained from independent nitrogen-sorption measurements (Figure S1). The narrower PSD observed in I07 reflects the sharper interfaces enabled by Ga intrusion.

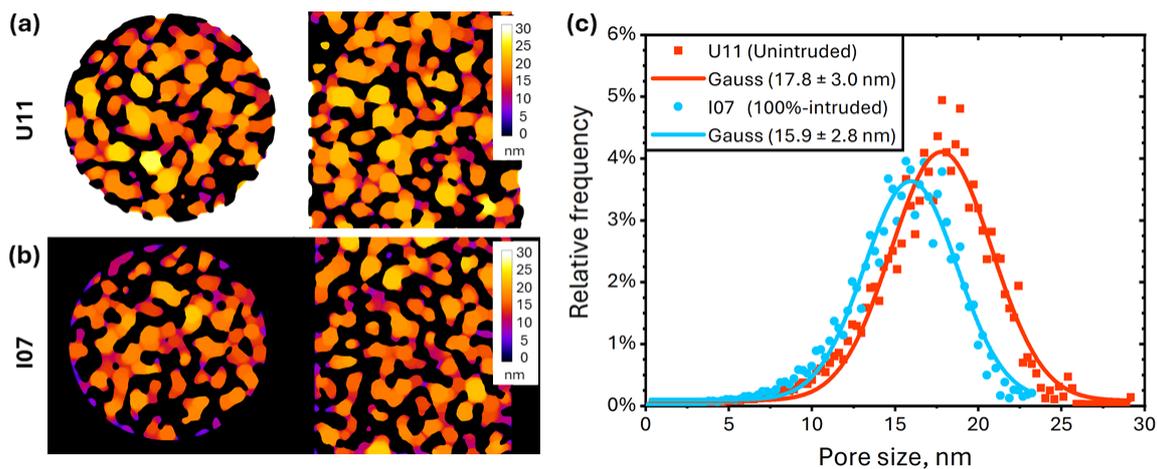

*Figure 6. Pore size distribution. Spheres of different sizes fitted into pores for (a) U11 and (b) I07. (c) The corresponding PSD data with Gaussian fitting.*

## 4. Discussion

The results of the study clearly demonstrate that Ga intrusion of mesoporous silica effectively overcomes two longstanding challenges in electron tomography of light-element non-conductive materials: i) low scattering contrast and ii) charging during sample preparation. In this work it was demonstrated that Ga intrusion significantly boosts HAADF-STEM imaging contrast and simplifies FIB preparation by improving the material's ability to dissipate charge, thereby enhancing the overall quality of 3D reconstructions and segmentation.

The BSE images reveal that Ga intrusion greatly enhances sample conductivity, thus preventing drift during image acquisition. During U11 pillar preparation, severe charging required extra pre-cutting and specialized charge-mitigation steps, while I07 and H02 were prepared with standard FIB lift-out procedures. The improved conductivity minimizes beam-induced artifacts during milling, ensuring uniform, defect-free pillars that are essential for obtaining high-fidelity tomography data.

The HAADF-STEM images confirm that Ga intrusion provides a significant increase in contrast. The 5.5-fold enhancement in HAADF intensity for Ga-intruded regions, as quantified by proxigram analysis, directly stems from the higher atomic number of Ga compared to silica. This improved contrast is crucial because it sharply delineates the Ga-intruded pores from both the silica matrix and carbon contamination, thereby reducing segmentation uncertainty. The ability to isolate the Ga phase in I07, and resolve three distinct phases in H02, has profound implications for accurate 3D reconstruction and subsequent quantitative analyses. Increased CL giving an additional 56% boost of intensity for the same electron dose makes the camera length another important parameter to consider for imaging of amorphous light-element beam-sensitive materials.

Quantitative assessments using FSC and proxigram methods indicate that the resolution is enhanced from 3.5 nm in unintruded samples to 2.3 nm in Ga-intruded samples. This resolution improvement is attributed not only to enhanced contrast, but also to the mechanical stabilization provided by Ga infiltration. The unintruded pillars exhibited deformation under electron beam exposure, whereas the Ga-intruded pillars maintained their structure throughout the tilt series. This mechanical stability is essential for reliable alignment during reconstruction and for achieving high-resolution imaging in ET.

Besides the significant improvements in imaging and resolution, the intrinsic pore structure of mesoporous silica remains unaffected by Ga infiltration. Analysis of both U11 and I07 shows similar values for tortuosity and connectivity, and the pore size distributions are consistent with independent nitrogen-sorption measurements. This finding is important because it confirms that the infiltration process enhances imaging without altering the material's inherent properties – a key requirement for accurate structure-property correlations. The narrower PSD in I07 reflects the sharper interface definition and enhanced resolution achieved through Ga intrusion.

The lower mean pore size in the 100%-intruded I07 sample may be partly attributed to the formation of a thin Ga oxide layer at the Ga-$SiO_2$ interface (see Supporting Information, Section S3). Although Ga does not typically react with $SiO_2$ under standard conditions, the presence of residual oxygen (originating either from the silica network or ambient trace gases) may promote limited oxidation of Ga. This can lead to the formation of Ga oxide near the pore walls, which exhibits a lower HAADF-STEM intensity compared to metallic Ga. Consequently, during segmentation, the lower intensity of the Ga oxide regions may result in an underestimation of the pore wall dimensions, which in turn leads to a lower apparent mean pore size in the I07 sample.

## 5. Conclusion and outlook

Our study shows that intruding mesoporous silica with Ga greatly improves the electron tomography workflow. By introducing Ga into the pore volume, we were able to reduce the charging problems during focused ion beam milling and increase the image contrast in HAADF-STEM. This led to clearer 3D reconstructions with a 34% improvement of the resolution in Ga-intruded specimens and resulted in a 49% sharper interface between the Ga and the silica matrix. The high contrast in 100%-intruded sample, which is five times stronger than in unintruded silica, significantly enhances phase differentiation during segmentation. This contrast improvement analogous to the effect of staining agents in medical imaging, which are used to highlight soft tissues. Importantly, the process of Ga-intrusion does not alter the geometric structure of the material, preserving the essential pore network properties, such as tortuosity and connectivity.

These findings suggest that Ga intrusion is a promising approach for addressing common challenges in imaging light-element beam-sensitive materials. Nonetheless, achieving uniform Ga infiltration in larger or more complex particles remains challenging, as the BSE imaging results indicate. Variations in pore size, the occurrence of trapped gas pockets, and differences in surface chemistry can lead to uneven Ga distribution. Future work should focus on optimizing pressure and temperature protocols and exploring surface treatments to improve the consistency of Ga infiltration. In addition, further optimization of imaging conditions – such as adjusting the camera length and electron dose – may lead to even better resolution and signal-to-noise ratios. The method presented here could also be extended to other porous materials, e.g., zeolites, organic–inorganic hybrids, or biomineralized structures that also suffer from low contrast and beam sensitivity. Adapting this technique to lab-based X-ray nano-tomography may allow high-contrast 3D imaging of larger sample volumes, making it useful for studying hierarchical pore systems.

Taken together, the use of Ga intrusion not only simplifies sample preparation and enhances imaging contrast but also enables more accurate 3D analyses and a deeper understanding of how pore structure relates to material properties in a broad range of functional porous materials.

## 6. Acknowledgements

The authors gratefully acknowledge the financial support by the German Research Foundation (DFG) within the frameworks of the Collaborative Research Centre 1411 (Project-ID 416229255) "Design of Particulate Products" and the Collaborative Research Centre 1452 (Project-ID 431791331) "Catalysis at Liquid Interfaces". Thanks to Colin Ophus from the National Center for Electron Microscopy (Berkeley, US) for providing custom apertures for STEM imaging. Thanks to Georg Haberfehlner from the Institute of Electron Microscopy and Nanoanalysis (TU Graz, Austria) for providing the "ToReAl" Matlab script for stack tilt series alignment.

## 7. CRediT authorship contribution statement

Alexander Kichigin: Writing - Original Draft, Investigation, Methodology, Formal analysis

Johannes Böhmer: Writing - Original Draft, Investigation, Formal analysis, Visualization

Moritz Buwen: Writing - Original Draft, Formal analysis, Visualization

Benjamin Apeleo Zubiri: Writing - Review & Editing, Methodology, Funding acquisition, Supervision

Mingjian Wu: Writing - Review & Editing, Methodology

Johannes Will: Writing - Review & Editing, Methodology

Dominik Drobek: Investigation

Alexander Götz: Investigation

Nora Vorlaufer: Writing - Review & Editing, Formal analysis

Jakob Söllner: Writing - Original Draft, Methodology, Resources

Matthias Thommes: Methodology, Resources, Supervision

Peter Felfer: Formal analysis, Supervision

Thomas Przybilla: Investigation, Writing - Review & Editing, Methodology, Supervision, Project administration

Erdmann Spiecker: Writing - Review & Editing, Project administration, Funding acquisition, Supervision, Methodology, Conceptualization

## 8. Declaration of competing interest

The authors declare that they have no known competing financial interests or personal relationships that could have appeared to influence the work reported in this paper.

## 9. Declaration of generative AI and AI-assisted technologies in the writing process

During the preparation of this work the authors used OpenAI ChatGPT in order to improve the language and readability. After using this service, the authors reviewed and edited the content as needed and take full responsibility for the content of the publication.

## 10. Data Availability Statement

Data for this paper, including all data included in the figures, is available at https://doi.org/10.5281/zenodo.14866207

# Supporting information for:

# Improving electron tomography of mesoporous silica by Ga intrusion

**Alexander Kichigin[a], Johannes Böhmer[a], Moritz Buwen[a], Benjamin Apeleo Zubiri[a], Mingjian Wu[a], Johannes Will[a], Dominik Drobek[a], Alexander Götz[a], Nora Vorlaufer[c], Jakob Söllner[b], Matthias Thommes[b], Peter Felfer[c], Thomas Przybilla[a]*, Erdmann Spiecker[a]***

[a] Institute of Micro- and Nanostructure Research (IMN) and Center for Nanoanalysis and Electron Microscopy (CENEM), Interdisciplinary Center for Nanostructured Films (IZNF), Department of Materials Science, FAU Erlangen-Nuremberg, Germany

[b] Institute of Separation Science and Technology, Department of Chemical and Biological Engineering, FAU Erlangen-Nuremberg, Germany

[c] Institute for General Materials Properties, Department of Materials Science, FAU Erlangen-Nuremberg, Germany

*Corresponding authors: thomas.przybilla@fau.de, erdmann.spiecker@fau.de

## S1. FD121 textural characterization.

Textural characterization was performed with high resolution nitrogen physisorption at 77 K using an Autosorb-iQ XR sorption instrument (Anton Paar QuantaTec, Boynton Beach, Florida). Nitrogen as adsorptive was used with a purity of 99.9999 % (Alphagaz2, Air Liquide). The CPG was outgassed prior to the experiment for 12 h at 150 °C under turbomolecular pump vacuum.

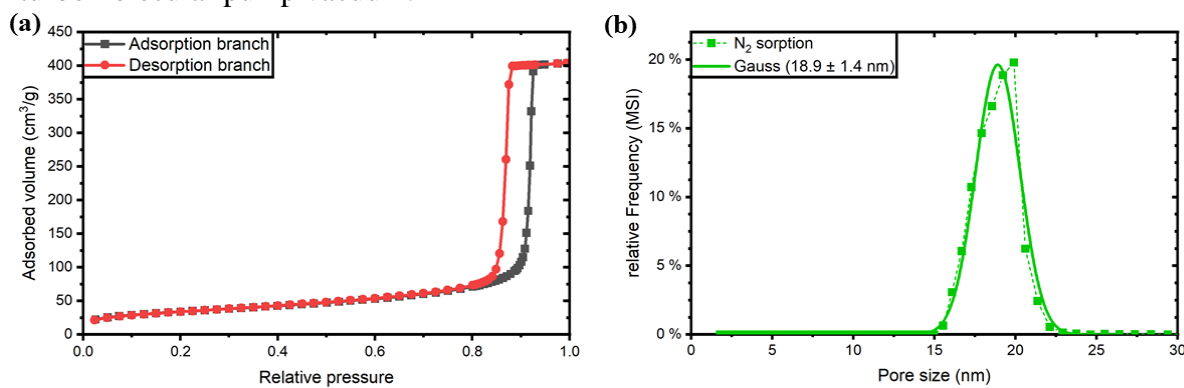

*Figure S1. (a) Sorption isotherm of the CPG and (b) pore size distribution of the FD121 mesoporous silica calculated from the isotherm and fitted with Gaussian distribution.*

## S2. Theoretical contrast calculations

To estimate the theoretical contrast enhancement in HAADF-STEM imaging due to Ga intrusion, we first determine the effective atomic number of silica ($SiO_2$) as

$$Z_{eff,SiO_2} = \frac{Z_{Si} + 2*Z_O}{3} = \frac{14 + 8*Z_O}{3} = 10$$

29  Ga, with an atomic number of 31, is expected to yield a higher HAADF signal. The
30  HAADF intensity is assumed to scale according to

$$I_{HAADF} = Z_{eff}{}^n$$

32  with the exponent nn typically in the range of 1.6 to 1.9. To further account for differences
33  in material density ($\rho$) and molar mass ($M$), the intensity is modeled as

$$I_{HAADF} = Z_{eff}{}^n * \frac{\rho}{M}$$

35  The contrast between two materials is then calculated as

$$C = \frac{\Delta I}{I_1} = \frac{I_2 - I_1}{I_1}$$

37  Using silica as the reference material (with its measured contrast set to 0.9 in unintruded
38  regions, since the background signal is always higher than zero), the predicted contrast
39  enhancement for Ga-intruded areas can be calculated by comparing the normalized
40  intensities. For example, with an exponent of n=1.7, the calculated theoretical contrast is 4.0,
41  whereas for n=2.0 (pure Rutherford scattering) the contrast is 5.9. When the predicted values
42  are normalized to the baseline contrast of 0.9, the resulting contrast improvements are 4.5x
43  for n=1.7 and 6.5x for n=2.0. Notably, at n=1.8 the model predicts a contrast enhancement
44  of approximately 5.1x, which is in excellent agreement with the experimentally observed
45  enhancement of about 5.5x.

## S3.  Ga oxide formation

For the investigation of the Ga–SiO$_2$ interface, a lamella of porous silica (with an average pore size of 150 nm) was analyzed using HAADF-STEM imaging and EDX mapping. An overview TEM image of the porous sample is shown in Figure S2. As can be observed, the Ga–SiO$_2$ interface is not always oriented edge-on to the electron beam. In particular, in the left part of the image, some Ga-intruded pores appear either in front of or behind a silica wall, resulting in a slightly darker gray appearance compared to the edge-on pores in the center. The red-marked area indicates the region where the EDX mapping (Figure S2b–d) was performed.

The complete EDX map was generated by integrating individual images acquired with a fast dwell time. During the measurement, changes in the sample were visible in the HAADF-STEM images. Figure S2b shows the first frame of the measurement, while Figure S2c displays the final frame (frame 2768). It is evident that the interface shifted (presumably to the right) which suggests that Ga flowed away; the first noticeable change occurred at the frame 1500. Inspection of the EDX map in Figure S2 reveals that the left side of the interface contains a mixture of silicon (blue) and oxygen (green), while the bright region, as expected, is predominantly Ga with only minor oxygen.

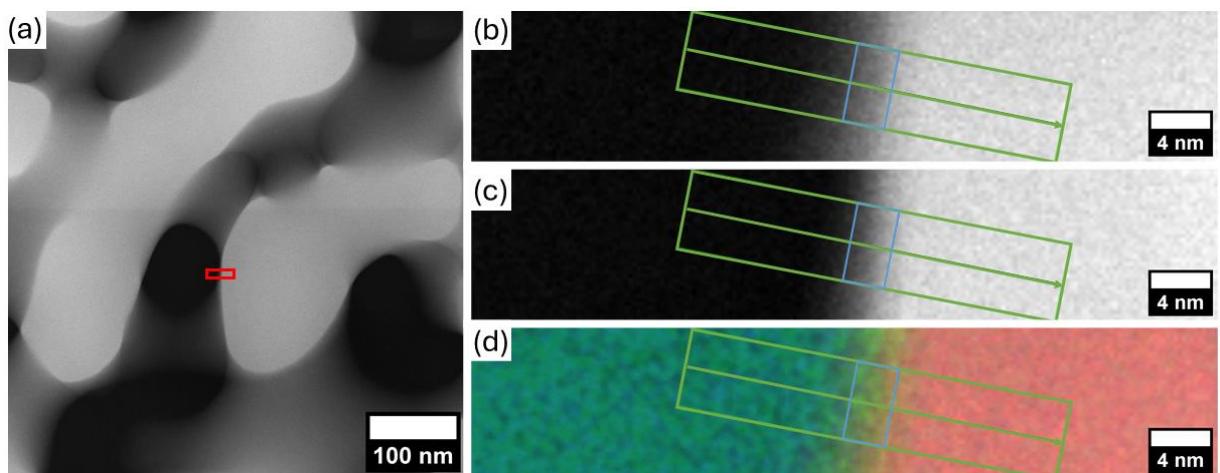

*Figure S2. (a) HAADF-STEM image of a lamella from the intruded silica with 150 nm pores. The red rectangle marks the region of interest used for EDX analysis (see b–d). (b) HAADF-STEM image at the beginning of the EDX analysis (Frame 1). (c) HAADF-STEM image at the end of the EDX analysis (Frame 2768), after more than 20 minutes of measurement. (d) Overlay of the net intensities of Ga (red), silica (blue), and oxygen (green). The green arrows indicate the ROI for the line profiles presented later. The green box represents the integration width, while the blue box (placed at the visible interface) serves as a reference.*

The green arrow in the mapping images indicates where the line profiles (Figure S3 and Figure S4) were extracted. The width of the green box corresponds to the integration width used for the line profiles, and the blue box serves as a reference marker. On the left side of the interface, the atomic percentages of oxygen and silicon match those expected for silicon dioxide. Moving to the right, both curves decrease – with silicon reaching 0% while oxygen remains at approximately 10%. The remainder of the right side consists of about 90% Ga, and as expected, no Ga is detected in the pore walls. At the 14 nm distance from the interface, where silicon is only 3%, the composition is approximately 59% Ga and 38% oxygen. Additionally, the HAADF detector signal at the interface closely follows the Ga profile, confirming that Ga is the primary contributor to the observed intensity.

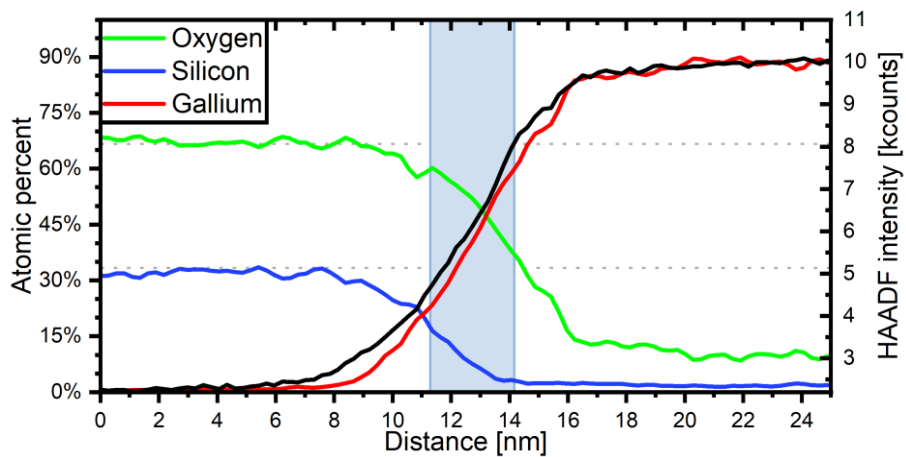

*Figure S3. Elemental distribution for oxygen, silicon, and Ga (along with the HAADF detector intensity) measured across the Ga–SiO$_2$ interface as indicated by the arrow in Figure 41.*

To further elucidate the behavior of oxygen at the interface, net intensities were analyzed using line profiles (Figure S4). It is evident that the silicon intensity drops off earlier than the oxygen intensity. By fitting a Boltzmann function to the curves, the distance between the 50% intensity points of the oxygen and silicon profiles was determined to be 2.7 nm. This indicates that oxygen extends 2.7 nm further into the interface than silicon – a value that corresponds to the width of the blue reference rectangle.

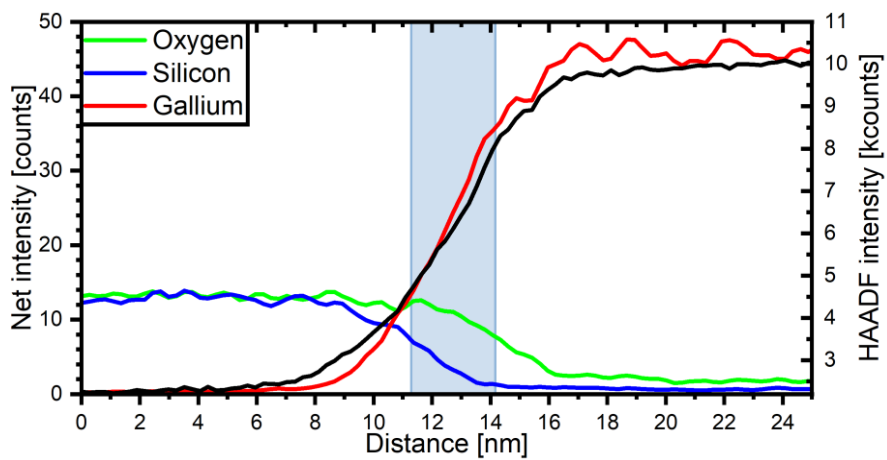

*Figure S4. Net intensity for oxygen, silica and Ga measured with EDX and the HAADF detector intensity across the Ga – silica interface as indicated with the arrow in Figure S2.*

## S4. LM images of the preparation of unintruded silica

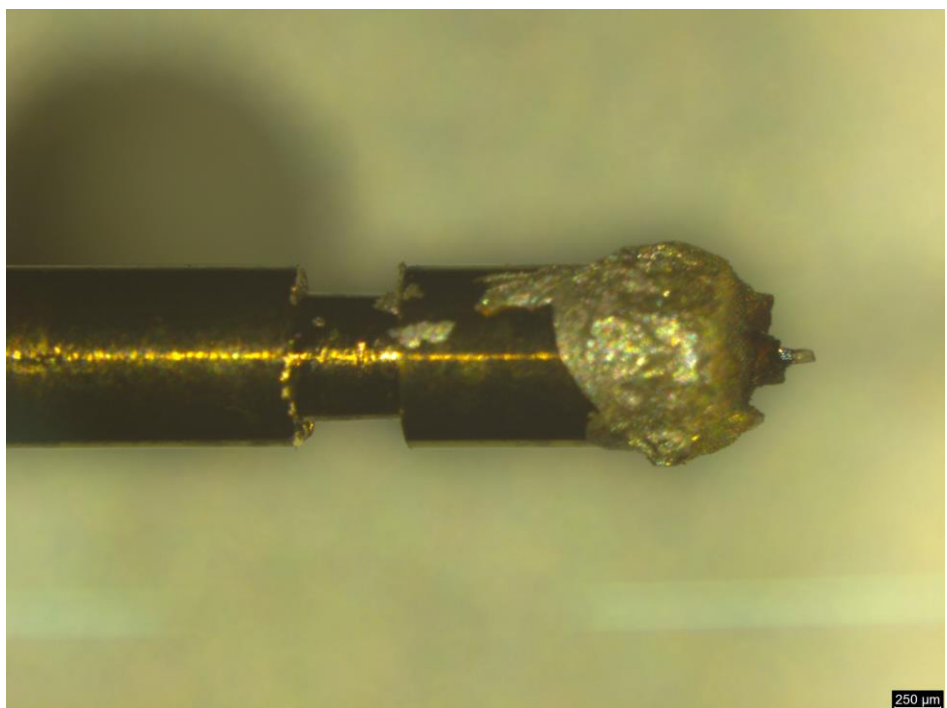

*Figure S5. The sample glued on the tip of the tomography pillar, thinned down with laser cutting.*

## S5. Deformation of unintruded pillars

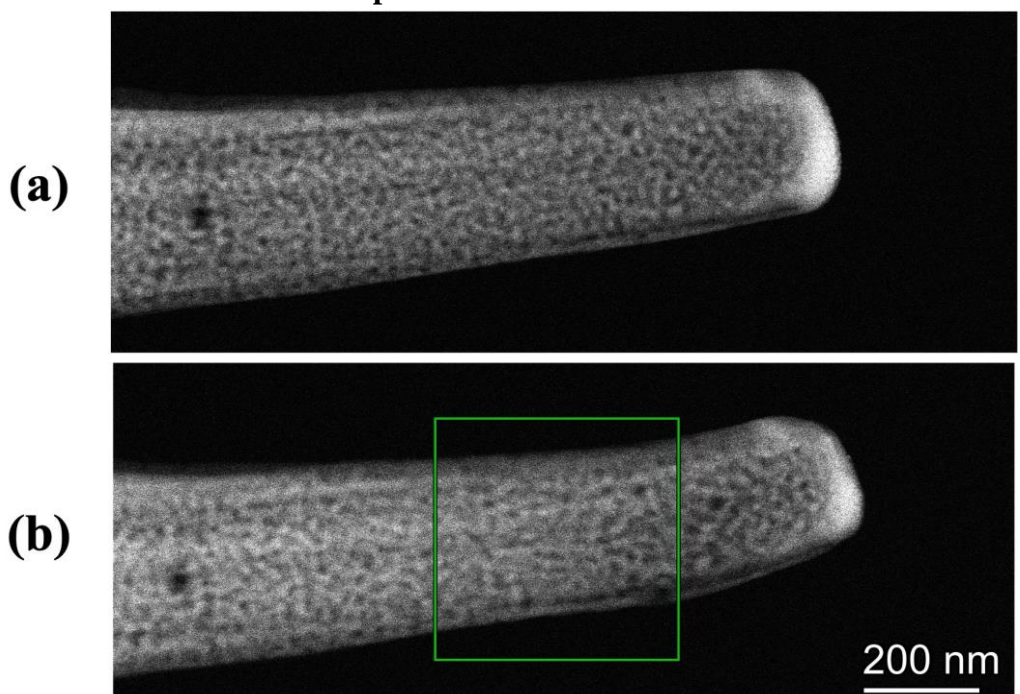

*Figure S6. The unintruded mesoporous silica before (a) and after (b) ET tomography of 360 projections taken with the 18 pA screen current, 0.27 nm pixel size, 5 us dwell time, resulting in the overall dose of 78 $e^-/Å^2$ per tilt. The green rectangle represents the imaging area, where the sample visibly deformed.*

## S6. Deformation of intruded pillars

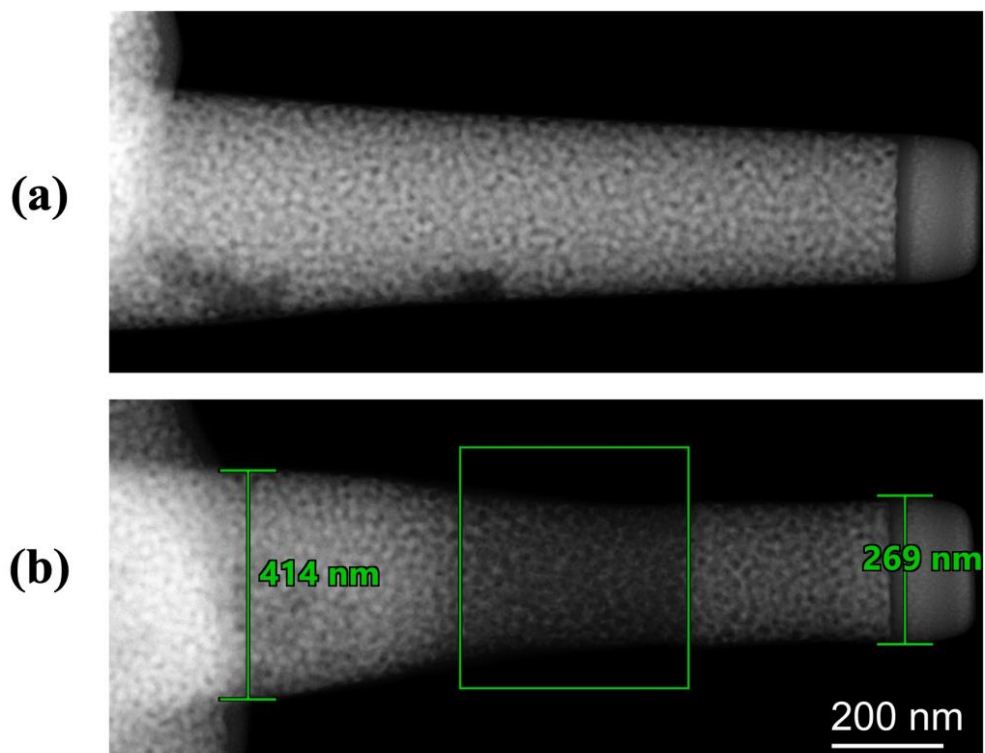

*Figure S7. The 100%-intruded mesoporous silica (a) before and (b) after ET tomography of 360 projections taken with the 60 pA screen current, 0.27 nm pixel size, 5 us dwell time. The green rectangle represents the imaging/focusing area. The sample visibly deformed in the imaging area.*

## S7. Detector image for quantification

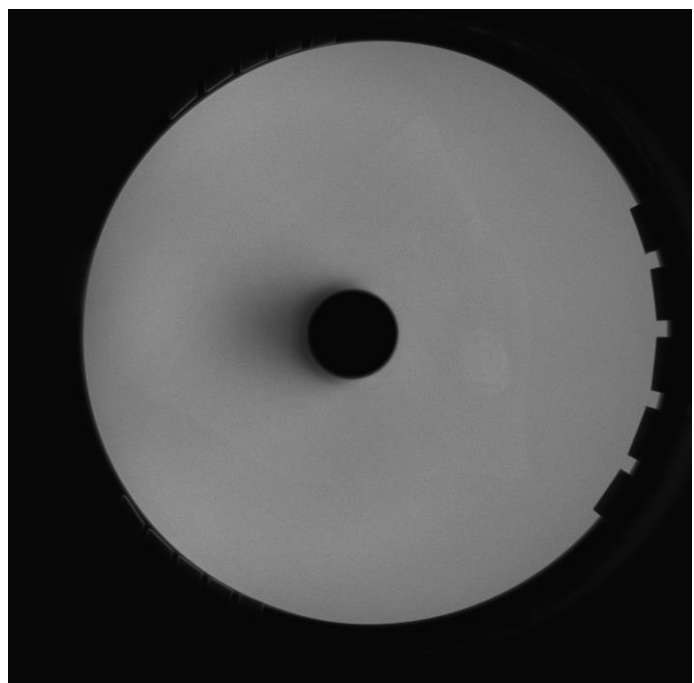

*Figure S8. The image of the HAADF detector used for quantifying the tilt series signal.*

## S8. Schematic of pore network analysis

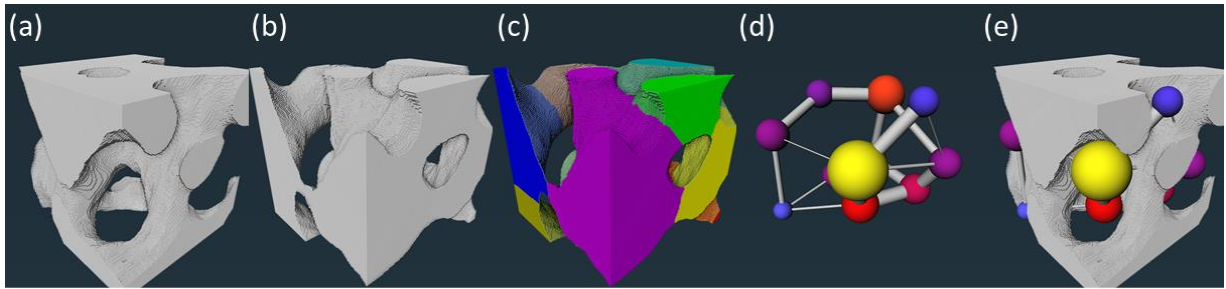

*Figure S9. Analysis of the pore network when using Amira Avizo to get the tortuosity and the connectivity of the network. (a) The segmentation of the pore walls is inverted to get in (b) the pores space. (c) Watershed algorithm separates the single pores and colors them differently only for visualization purpose. (d) Visualization of the pores by balls and the throats by sticks connecting them. The color and the size correspond to the equivalent pore radius. The thickness of the throats corresponds to the throat radius. (e) Modeled pore network inside the pore walls.*

## S9. Histograms of the reconstructions

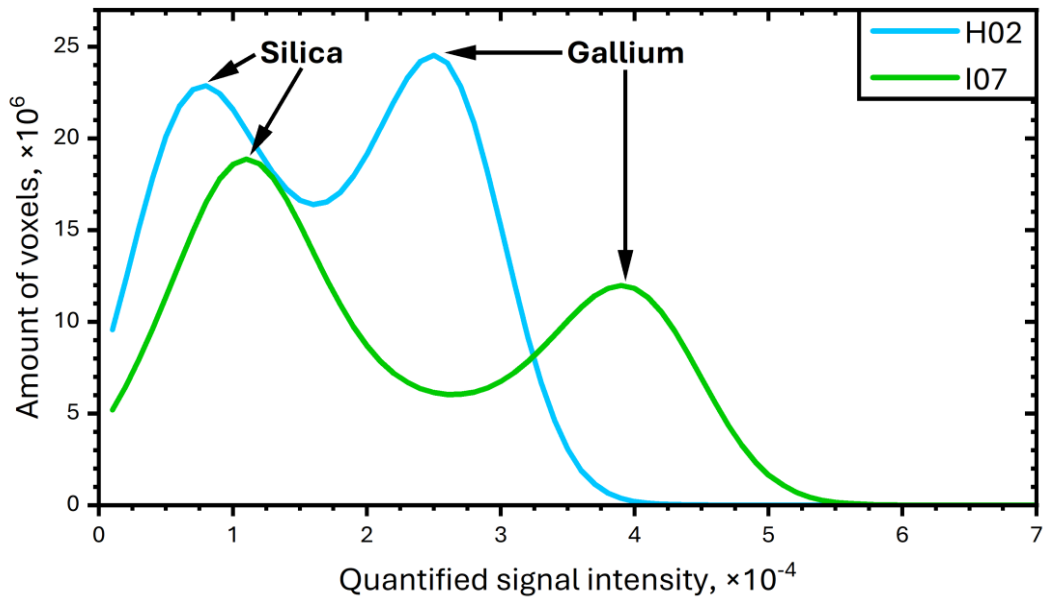

*Figure S10. Histogram analysis of the 3D reconstructions for 100%-intruded I07 and 50%-intruded H02. The main experimental difference is that I07 was taken with the Camera length of 91 mm, while H02 – 185 mm. Higher Camera length expectably leads to higher intensity of the HAADF signal and better separation between two phases (Ga and silica). The left peak on both plots corresponds to the signal of the silica and is improved from $8 \cdot 10^{-5}$ to $11 \cdot 10^{-5}$ (37% improvement), while the right-peak Ga signal is improved from $2.5 \cdot 10^{-4}$ to $3.9 \cdot 10^{-4}$ (56% improvement).*

## S10. Connectivity distribution

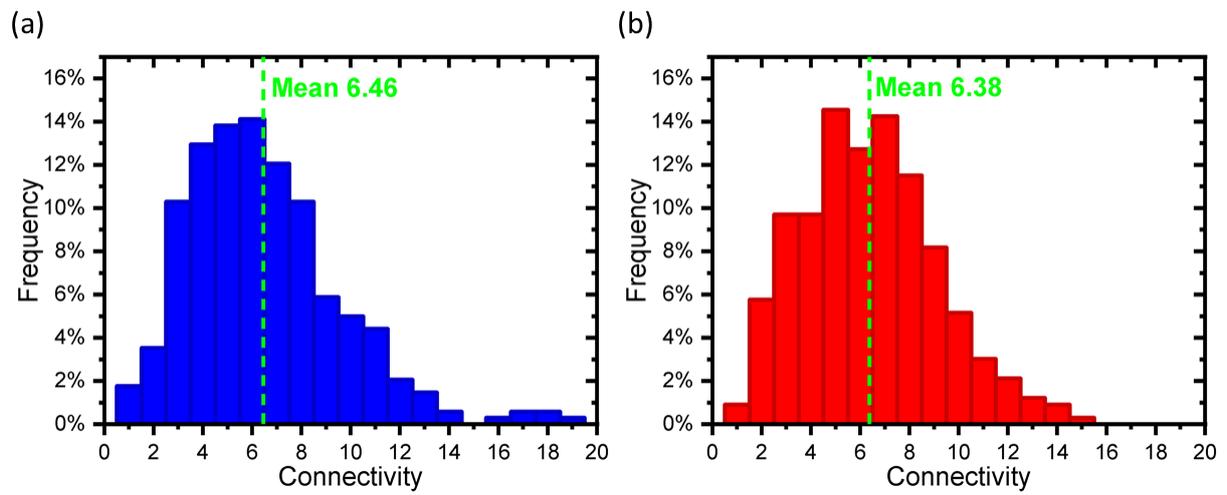

*Figure S11. Connectivity distribution calculated from the porous silica for (a) the unintruded and (b) the 100%-intruded sample.*